\documentclass[prc,twocolumn,nofootinbib,showpacs,amsmath,amssymb]{revtex4-1}
\usepackage{dcolumn}
\usepackage{bm}
\usepackage{color}
\usepackage{amssymb}
\usepackage{amsmath}
\usepackage{graphicx}
\usepackage{amsfonts}
\usepackage{slashed}
\usepackage{pstricks}
\usepackage{float}
\usepackage{hyperref}
\usepackage{array}
\allowdisplaybreaks
\usepackage{dcolumn}
\usepackage{epsf}
\usepackage{isotope}
\begin{document}
\title{Extracting Hypernuclear Properties from the $(e, e^\prime K^+)$ Cross Section}
\author{Omar Benhar}
\email{omar.benhar@roma1.infn.it}
\affiliation{INFN and Department of Physics, ``Sapienza'' University, I-00185 Rome, Italy}

\date{\today}

\begin{abstract}
Experimental studies of hypernuclear dynamics, besides being essential for the understanding of
strong interactions in the strange sector, have important astrophysical implications. The observation
of neutron stars with masses exceeding two solar masses poses a serious challenge to the
models of hyperon dynamics in dense nuclear matter, many of which predict a maximum mass
incompatible with the data.
In this paper, it is argued that valuable new insight can be gained from the forthcoming extension of the experimental studies
of  kaon electro production from nuclei to include the $\isotope[208][]{\rm Pb}(e,e^\prime K^+) \isotope[208][\Lambda]{\rm Tl}$ process.
A comprehensive framework for the description of kaon electro production, based on factorisation of the nuclear cross section and the formalism of nuclear many-body theory, is outlined. This approach highlights the connection between the kaon production and proton knockout reactions, which will allow to exploit  the available $\isotope[208][]{\rm Pb}(e,e^\prime p) \isotope[207][]{\rm Tl}$ data to achieve  a largely model-independent analysis of the measured cross section.
\end{abstract} 


\index{}\maketitle

\section{Introduction}
\label{intro}

Experimental studies of the $(e,e^\prime K^+)$ reaction on nuclei  have long been recognized as a valuable
source of information on hypernuclear spectroscopy. The extensive program of measurements
performed or approved at Jefferson Lab~\cite{E94-107,E12-15-008}\textemdash encompassing a variety of nuclear targets
ranging from $\isotope[6][]{\rm Li}$ to $\isotope[40][]{\rm Ca}$ and $\isotope[48][]{\rm Ca}$\textemdash
has the potential to shed new light on the dynamics of strong interactions in the strange sector, addressing
outstanding issues such as the isospin-dependence of hyperon-nucleon interactions  and the role of three-body forces involving nucleons and hyperons. In addition, because the appearance of hyperons is expected to become energetically favoured in dense nuclear matter, these measurements have important implications for neutron star physics.

The recent observation of two-solar-mass neutron stars~\cite{demorest,antonio}\textemdash the existence of which is ruled out by many
models predicting the presence of hyperons in the neutron star core~\cite{isaac_etal}\textemdash  suggests that
the present understanding of nuclear interactions involving hyperons is far from being complete.
In the literature, the issue of reconciling the calculated properties of hyperon matter with the existence of massive stars is referred to as
{\em hyperon puzzle}~\cite{puzzle}.

Owing to the severe difficulties involved
in the determination of the potential describing hyperon-nucleon (YN) interactions from scattering data, the study of hypernuclear spectroscopy
provides an effective alternative approach, capable of yielding much needed additional information.

In this context, the $(e,e^\prime K^+)$ process offers clear advantages.
The high resolution achievable by $\gamma$-ray spectroscopy can only be exploited to study energy levels
below nucleon emission threshold, while $(K^-,\pi^-)$ and $(\pi^+, K^+)$ reactions mainly provide
information on non-spin-flip interactions. Moreover, compared to hadron induced reactions, kaon electro production
allows a better energy resolution, which may in turn result in a more accurate identification of the hyperon binding energies~\cite{E94-107}.
However, the results of several decades of study of the $(e,e^\prime p)$ reaction show that,
to achieve this goal, the analysis of the measured cross sections must be based on a theoretical model taking into account
the full complexity of electron-nucleus interactions~\cite{Benhar:NPN}.
Addressing this issue will be  critical to the recently approved extension of the Jefferson Lab program to the case of a heavy target with large neutron
excess, the nucleus $\isotope[208][]{\rm Pb}$. This experiment will allow to study hyperon dynamics in an environment providing
the best available proxy of the neutron star interior. A short account of the proposal to measure the
$\isotope[208][]{\rm Pb}(e,e^\prime K^+) \isotope[208][\Lambda]{\rm Tl}$ cross section in Jefferson Lab Hall A,
can be found in Ref.~\cite{franco_AIP}.

This work is meant to lay down the foundations of a comprehensive theoretical  framework for the description  of the
 $(e,e^\prime K^+)$ cross section within the formalism of nuclear many-body theory, which has been extensively and successfully employed to study the
proton knockout reaction~\cite{Benhar:NPN}. In addition, it is shown that, owing to the connection between $(e,e^\prime p)$  and
$(e,e^\prime K^+)$ processes which naturally emerges in the context of the proposed analysis, the missing energy
spectra measured in $(e,e^\prime p)$ experiments provide the baseline needed for a largely model-independent determination
of the hyperon binding energies.

The body of the paper is structured as follows. In Sect.~\ref{Axsec} the treatment of kaon electro-production
from nuclei in the kinematical regime in which factorisation of the nuclear cross section is expected to be applicable is derived,
and the relation to the proton knockout process is highlighted. The main conceptual issues associated with the description of the elementary
electron-proton vertex and the nuclear amplitudes comprising the structure of the
$\isotope[208][]{\rm Pb}(e,e^\prime K^+) \isotope[208][\Lambda]{\rm  Tl}$ cross section are discussed in Sect.~\ref{Pbxsec}.
Finally,  the summary and an outlook to future developments can be found in in Sect.~\ref{summary}.

\section{The ${\rm A}(e, e^\prime K^+){_\Lambda}{\rm A}$ cross section}
\label{Axsec}

Let us consider the kaon electro-production process
\begin{equation}
\label{eek:A}
e(k) + {\rm A}(p_{\rm A}) \to e^\prime(k^\prime) + K^+(p_K) + {_\Lambda}{\rm A}(p_R) \ ,
\end{equation}
in which an electron scatters off a nucleus of mass number ${\rm A}$, and the hadronic final state
\begin{equation}
\label{def:F}
| F \rangle = | K^+ {_\Lambda}{{\rm A}} \rangle \ ,
\end{equation}
includes a $K^+$ meson
and the recoiling
hypernucleus, resulting from the replacement of a proton with a $\Lambda$ in the target nucleus.
The incoming and scattered electrons have four-momenta $k \equiv (E,{\bf k})$ and $k^\prime \equiv(E^\prime,{\bf k}^\prime)$,
respectively, while the corresponding quantities associated with  the kaon and the recoiling hypernucleus are denoted   $p_K \equiv (E_K,{\bf p}_k)$ and
$p_R \equiv(E_R,{\bf p}_R)$. Finally, in the lab reference frame\textemdash in which the lepton kinematical variables are measured\textemdash
$p_A \equiv(M_A,0)$.

The differential cross section of reaction (\ref{eek:A}) can be written in the form
\begin{equation}
\label{A:xsec}
d \sigma_A \propto L_{\mu\nu} W^{\mu\nu}  \ \delta^{(4)}( p_A + q - p_F) \ ,
\end{equation}
with $\lambda, \mu = 1,2,3$, where $q = k - k^\prime$ and $p_F~=~p_K + p_R$ are the four-momentum transfer and the total four-momentum carried by the hadronic final state, respectively. The tensor $L_{\mu\nu}$, fully
specified by the electron kinematical variables, can be written in the form \cite{AFF}

\begin{equation}
L = \left(
\begin{array}{ccc}
\eta_+ & 0 & -\sqrt{\epsilon_L \eta_+} \\
0      & \eta_- & 0 \\
-\sqrt{\epsilon_L \eta_+} & 0 & \epsilon_L \\
\end{array}
\right) \ ,
\end{equation}
with $\eta_\pm = \left( 1 \pm \epsilon \right)/2$  and
\begin{equation}
\epsilon = \left( 1 + 2 \frac{|{\bf q}|^2}{Q^2}\ \tan^2 \frac{\theta_e}{2} \right)^{-1}  \ \ ,
\end{equation}
where $\theta_e$ is the electron scattering angle, $q \equiv ( \omega, {\bf q} )$, $Q^2 = - q^2$,  and
$\epsilon_L =  \epsilon Q^2 / \omega^2$.

All the information
on hadronic,  nuclear and hypernuclear dynamics is contained in the nuclear response tensor, defined as
\begin{equation}
\label{A:tensor}
W^{\mu\nu} = \langle 0 | {J_{\rm A}^\mu}^\dagger(q) | F \rangle \langle F | J_{\rm A}^\nu(q) | 0 \rangle  \ ,
\end{equation}
where $|0 \rangle$ denotes the target ground state and the final state $|F\rangle$ is given by Eq.~(\ref{def:F}).

Equation (\ref{A:tensor}) shows that the theoretical calculation of the cross section
requires
a consistent description of the nuclear and hypernuclear wave functions, as well as of the nuclear current operator
appearing in the transition matrix element, $ J_{\rm A}^\mu$.  This problem, which in general involves prohibitive
difficulties, greatly simplifies in the kinematical region in which the impulse approximation can be exploited.

\subsection{Impulse approximation and factorisation}
\label{IA}

\begin{figure}[h!]
\vspace*{-.1in}
\begin{center}
\includegraphics[scale=1.20]{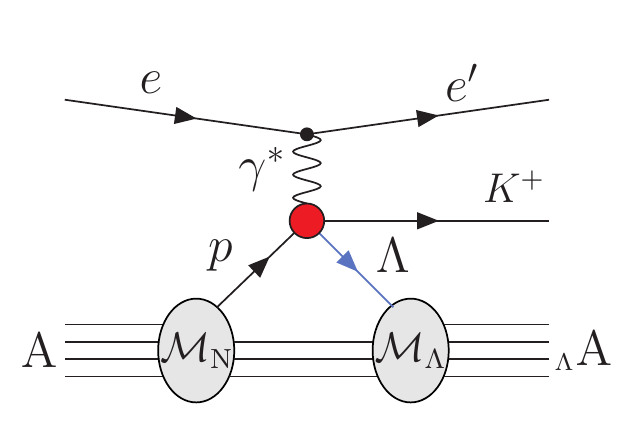}
\end{center}
\vspace*{-.2in}
\caption{Schematic representation of the scattering amplitude associated with the process of
Eq.~(\ref{eek:A})  in the impulse approximation regime.}
\label{graph}
\end{figure}

Figure~\ref{graph} provides a diagrammatic representation of the $(e,e^\prime K^+)$ process based on the
factorisation {\em ansatz}.
The formalism exploiting factorisation of the nuclear cross section is expected to be applicable in the impulse
approximation regime, corresponding to momentum transfer such that the wavelength of the virtual photon,
$\lambda~\sim~1/|{\bf q}|$, is short compared to the average distance between nucleons in the target nucleus,
$d_{NN}~\sim~1.5 \  {\rm fm}$.

Under these condition, which can be  easily met at Jefferson Lab\textemdash hereafter JLab\textemdash the beam particles primarily interact with individual protons, the remaining
${\rm A}-1$ nucleons acting as spectators. As a consequence,
the nuclear current operator reduces to the sum of one-body operators describing the electron-proton interaction
\begin{equation}
J_{\rm A}^\mu(q) = \sum_{i=1}^A j^\mu_i(q)\ ,
\end{equation}
and the hadronic final state takes the product form
\begin{equation}
| F \rangle = | K^+ \rangle \otimes \vert {_\Lambda}{{\rm A}} \rangle \ ,
\end{equation}
with the outgoing $K^+$ being described by a plane wave, or by a distorted wave obtained from a kaon-nucleus
optical potential~\cite{E94-107}.

From the above equations, it follows that the nuclear transition amplitude
\begin{equation}
{\mathcal M}^\mu = \langle K^+ {_\Lambda}{{\rm A}} | J_{\rm A}^\mu(q) | 0 \rangle \ ,
\end{equation}
can be written in factorized form
through insertion of the completeness relations
\begin{equation}
\int \frac{d^3p}{(2\pi)^3} | {\bf p} \rangle \langle {\bf p} |  = \int \frac{d^3p_\Lambda}{(2\pi)^3} | {\bf p}_\Lambda \rangle \langle {\bf p}_\Lambda |
=  \openone \ ,
\end{equation}
where the integrations over the momenta carried by the proton and the $\Lambda$ also include spin summations, and
\begin{equation}
\sum_n | ({\rm A}-1)_n \rangle \langle ({\rm A}-1)_n | =  \openone ,
\end{equation}
the sum being extended to all eigenstates of the $({\rm A}-1)$-nucleon spectator system.


The resulting expression turns out to be
\begin{widetext}
\begin{align}
\label{factorization}
\mathcal{M}^\mu
=  \sum_{i=1}^{\rm A} \sum_n \int \frac{d^3p}{(2\pi)^3} \frac{d^3p_\Lambda}{(2 \pi)^3}
{\mathcal M}^\star_{ _\Lambda {\rm A} \to ({\rm A}-1)_n + \Lambda}  \  \langle {\bf p}_K {\bf k}_\Lambda | j_i^\mu | {\bf p} \rangle
\ {\mathcal M}_{{\rm A} \to ({\rm A}-1)_n + p} \ \ ,
\end{align}
\end{widetext}
where the current matrix element describes the elementary electromagnetic process $\gamma^* + p \to K^+ + \Lambda$.

Note that, compared to the case of nucleon knock out discussed in, e.g.,  Ref.~\cite{benhar_RMP}, the transition
amplitude describing kaon production within the factorisation scheme, Eq.~(\ref{factorization}), exhibits a more complex structure,
involving two different amplitudes.

The nuclear and hypernuclear amplitudes in the right-hand side of Eq.~(\ref{factorization}), labelled  ${\mathcal M}_N$ and
${\mathcal M}_\Lambda$ in Fig.~\ref{graph}, are given by
\begin{equation}
\label{ampl:N}
{\mathcal M}_{{\rm A} \to ({\rm A}-1)_n + p} = \{ \langle {\bf p} | \otimes \langle ({\rm A}-1)_n | \} | 0 \rangle \ ,
\end{equation}
and
\begin{equation}
\label{ampl:Y}
{\mathcal M}_{_\Lambda {\rm A} \to ({\rm A}-1)_n + \Lambda}  = \{ \langle {\bf p}_\Lambda | \otimes \langle ({\rm A}-1)_n | \} | _\Lambda{{\rm A}} \rangle \ .
\end{equation}
In the above equations, the states $ \vert ({\rm A}-1)_n \rangle$ and $ \vert _\Lambda{\rm A} \rangle$ describe the $({\rm A}-1)$-nucleon spectator system, appearing as an intermediate state, and
the final-state $\Lambda$-hypernucleus, respectively.

The amplitudes of Eq.~(\ref{ampl:N}) determine the Green's function  $G({\bf k},E)$\textemdash embodying all information on single-particle dynamics in the target nucleus\textemdash and the associated spectral function, defined as
\begin{align}
\label{SF:N}
P({\bf k},E) & =   \frac{1}{\pi} {\rm Im} \ G({\bf k},E) \\
\nonumber
& = \sum_n  \vert {\mathcal M}_{{\rm A }\to ({\rm A}-1)_n + p}  \vert^2
 \ \delta(E + M_A-m-E_n) \ ,
\end{align}
where $m$ is the nucleon mass and $E_n$ denotes the energy of the $({\rm A}-1)$-nucleon system in the state $n$. The spectral function
describes the {\em joint} probability to remove a nucleon of momentum ${\bf k}$ from the nuclear ground state
leaving the residual system with excitation energy $E>0$.

Note that within the mean-field approximation, underlying the nuclear shell model,
Eq.~(\ref{SF:N}) reduces to the simple form
\begin{equation}
\label{SF:N:MF}
P({\bf k},E) = \sum_{\alpha \in \{F\}} |\varphi_\alpha({\bf k})|^2 \delta(E -  |\epsilon_\alpha|) \ ,
\end{equation}
where $\alpha \equiv \{ nj\ell \}$ is the set of quantum numbers specifying single-nucleon orbits. The sum is extended to
all states belonging to the Fermi sea, the momentum-space wave functions
and energies of which are denoted $\varphi_\alpha({\bf k})$ and $\epsilon_\alpha$, respectively, with $\epsilon_\alpha<0$.

Equation (\ref{SF:N:MF}) shows that within the independent particle model the spectral function reduces to a set
of $\delta$-function peaks, representing the energy spectrum of single-nucleon states. Dynamical effects
beyond the mean field shift the position of the peaks, that also acquire a finite width.
In addition, the occurrence of virtual scattering processes\textemdash bringing about the excitation of nucleon pairs to continuum states
above the Fermi surface\textemdash leads to the appearance of a sizeable smooth contribution to the Green's funcion,
accounting for $\sim 20 \%$ of the total strength. As a consequence, the normalisation of a shell model state
$\varphi_\alpha$, referred to as spectroscopic factor, is reduced from unity to a value $Z_\alpha <1$.

The nuclear spectral functions  have been extensively studied measuring the cross section of the $(e,e^\prime p)$ reaction,
in which the scattered electron and the knocked out nucleon are detected in coincidence.
The results of these experiments, carried out using a variety of nuclear targets, have
unambiguous identified the states predicted by the shell model, highlighting at the same time the limitations of the
mean-field approximation and the effects of nucleon-nucleon
correlations~\cite{Benhar:NPN,FruMug,eep_nikhef}.

In analogy with Eqs.~(\ref{ampl:N}) and (\ref{SF:N}), the amplitudes of Eq.~(\ref{ampl:Y}) comprise the spectral function
\begin{align}
\label{SF:L}
P_\Lambda({\bf k}_\Lambda,E_\Lambda) & =  \sum_n  \vert {\mathcal M}_{_\Lambda A \to (A-1)_n + \Lambda}  \vert^2 \\
& \times  \delta(E_\Lambda+ M_{_\Lambda {\rm A}} - M_\Lambda - E_n) \ ,
\end{align}
describing the joint probability to remove a $\Lambda$ from the hypernucleus $_\Lambda{\rm A}$ leaving the residual system with
energy $E_\Lambda$. Here $M_\Lambda$ and $M_{_\Lambda {\rm A}}$
denote the mass of the $\Lambda$ and the hypernucleus, respectively.


The observed $(e,e^\prime K^+)$ cross section, plotted as a function of the missing energy
\begin{equation}
\label{def:emiss}
E^\Lambda_{\rm miss} = \omega - E_{K^+} \ ,
\end{equation}
exhibits a collection of peaks, providing the sought-after information on the energy spectrum of the $\Lambda$ in the
final state hypernucleus\footnote{In principle, the right-hand side of Eq.~(\ref{def:emiss}) should also include a term accounting for the
kinetic energy of the recoiling hypernucleus. However, for heavy targets this contribution turns out to be negligibly small, and will be omited.} .
Note that both the electron energy loss, $\omega$, and the energy of the outgoing kaon,$E_{K^+}$, are {\em measured} kinematical quantities.

\subsection{Kinematics}
\label{Kin}

The expression of $E^\Lambda_{\rm miss}$, Eq.~(\ref{def:emiss}), can be conveniently rewritten considering that the $\delta$-function
of Eq.~(\ref{A:xsec}) implies  the condition
\begin{equation}
\label{full:encons}
\omega + M_A = E_{K^+} + E_{_\Lambda{\rm A}} \ .
\end{equation}
Combining the above relation with the requirement of conservation of energy at the nuclear and
hypernuclear vertices, dictating
\begin{equation}
\label{cons:ampl}
M_A = E_p + E_n \ \ \ , \ \ \ E_\Lambda + E_n = E_{_\Lambda{\rm A}} \ ,
\end{equation}
we find
\begin{equation}
\label{cons:vert}
\omega + E_p = E_{K^+} + E_\Lambda \  .
\end{equation}
Finally, substitution into Eq~(\ref{def:emiss}) yields
\begin{equation}
\label{lambda:emiss}
E^\Lambda_{\rm miss} = E_\Lambda - E_p \ .
\end{equation}

The above equation, while providing a relation between the {\em measured} missing energy and the binding energy of the $\Lambda$ in the
final state hypernucleus, defined as $B_\Lambda~=~-E_\Lambda$, {\em does not} allow for a model-independent identification of $E_\Lambda$.
The position of a peak observed  in the missing energy spectrum turns out to be determined by the difference between the energies
needed to remove a $\Lambda$ from the final state hypernucleus,  $E_\Lambda$, or a proton from the target nucleus, $E_p$,
leaving the residual $(A-1)$-nucleon system in the {\em same} bound state, specified by the quantum numbers collectively denoted $n$.

The proton removal energies, however, can be {\em independently} obtained from the missing energy {\em measured} by  proton
knockout  experiments\textemdash in which the scattered electron and the ejected proton are detected in coincidence\textemdash
defined as
\begin{equation}
\label{def:emissp}
E^p_{\rm miss} = \omega - E_{p^\prime} = - E_p \ .
\end{equation}
where $E_{p^\prime}$ is the energy of the outgoing proton. Note that,  consistently with Eq.~(\ref{def:emiss}), in the right-hand side of the above equation the kinetic energy of the recoiling nucleus has been omitted.

From Eqs.~(\ref{lambda:emiss}) and (\ref{def:emissp}) it follows that the $\Lambda$ binding energy can be determined in a fully model-independent fashion from
\begin{equation}
B_\Lambda = - E_\Lambda = - ( E^\Lambda_{\rm miss} - E^p_{\rm miss} ) \ ,
\end{equation}
combining the information provided by the missing energy spectra measured in $(e,e^\prime K^+)$ and $(e,e^\prime p)$ experiments.

\section{The
$\isotope[208][]{\rm Pb}(e, e^\prime K^+)\isotope[208][\Lambda]{\rm Tl}$ Cross Section}
\label{Pbxsec}

In view of astrophysical applications, the recently approved measurement of the
$\isotope[208][]{\rm Pb}(e, e^\prime K^+)\isotope[208][\Lambda]{\rm Tl}$ cross section at JLab~\cite{franco_AIP}
will be of outmost importance.
In this section, I will review the basic elements of the theoretical description of kaon electro production within the factorisation
scheme laid down in Sect.~\ref{Axsec}.

\subsection{The $e+p \to e^\prime + K^+ + \Lambda$ process}

The description of the elementary  $e+p \to e^\prime + K^+ + \Lambda$ process involving an isolated  proton at rest
has been obtained from the isobar model~\cite{Adam,isobar_model}, in which the hadron current is derived from an effective Lagrangian comprising baryon and meson fields. Different implementations of this model are characterised by the intermediate states appearing in
processes featuring the excitation of resonances~\cite{Sotona,petr1,petr2}. The resulting expressions\textemdash involving a set
of free parameters determined by fitting the available experimental data\textemdash have been used to obtain
nuclear cross sections within the approach based on the nuclear shell model and the frozen-nucleon
approximation~\cite{Sotona,E94-107}

In principle, the calculation of the nuclear cross section within the scheme outlined in Sect.~\ref{IA}  should  be performed
taking into account that the elementary process involves a bound, moving nucleon, with  four-momentum
$p \equiv (E_p,{\bf p})$ and energy
\begin{equation}
\label{offshell:momentum}
E_p = M_A - E_n = m - E \ ,
\end{equation}
as prescribed by Eq.~(\ref{SF:N}). However, the generalisation to off-shell kinematics of phenomenological approaches
constrained by free proton data, such as the isobar model of Refs.~\cite{Sotona,petr1,petr2}, entails non trivial conceptual difficulties.

A simple procedure to overcome this problem is based on the observation that in the scattering
process on a bound nucleon, a fraction $\delta \omega$ of the energy transfer goes to the spectator system.
The amount of energy given to the struck proton, the expression of which naturally emerges from the impulse approximation formalism, turns
out to be~\cite{benhar_RMP}
\begin{equation}
\label{omegatilde}
{\widetilde \omega} =  \omega -  \delta \omega = \omega + m - E - \sqrt{ m^2 + {\bf p}^2 }  \ .
\end{equation}
Note that from the above equations it follows that
\begin{equation}
\label{omegatilde2}
E_p + \omega  = \sqrt{ m^2 + {\bf p}^2 } + {\widetilde \omega}  \ ,
\end{equation}
implying in turn
\begin{equation}
\label{omegatilde2}
(p + q )^2 =  ( {\widetilde p} +  {\widetilde q} )^2  = W^2\ ,
\end{equation}
with
${\widetilde q} \equiv ( {\widetilde \omega} , {\bf q})$ and ${\widetilde p} \equiv ( \sqrt{ m^2 + {\bf p}^2 }, {\bf p})$.

The above equations show that the replacement $q \to {\widetilde q}$ allows to establish a correspondence
between electron scattering on an {\em off-shell} moving proton, leading to the appearance of a final state of invariant mass $W$, and
the corresponding process involving a proton {\em in free space}. As a consequence, within this scheme the use of an
elementary cross section designed to explain hydrogen data is fully consistent.

It has to be mentioned that, although quite reasonable on physics grounds, the use of ${\widetilde q}$ in the hadron current leads
to a violation of current conservation. This issue is inherent in the impulse approximation scheme,
which does not allow to simultaneously conserve energy and current in correlated systems.
A very popular and effective workaround for this problem, widely employed in the analysis of $(e,e^\prime p)$ data,
has been first proposed by de Forest in the 1980s~\cite{forest}.

In view of the fact that the extension of the work of Refs.\cite{petr1,petr2} to the case of a moving proton does not
involve severe conceptual difficulties, the consistent application of the formalism developed for
proton knock-out processes to the case of kaon electro production appears to be feasible.
In this context, it should also be pointed out that the factorisation scheme discussed in Sect.~\ref{Axsec}
allows for a fully relativistic treatment of the electron-proton vertex, which is definitely required in the kinematical
region accessible at JLab~\cite{benhar_RMP}.

\subsection{Nuclear and Hypernuclear Dynamics}

Valuable information needed to obtain $\Lambda$ removal energies from the cross section of the
$\isotope[208][]{Pb}(e, e^\prime K^+)\isotope[208][\Lambda]{Tl}$ process, using the procedure described in Sect.~\ref{Kin}, has been gained by the high-resolution studies of the
$\isotope[208][]{Pb}(e,e^\prime p)\isotope[207][]{Tl}$ reaction performed at
NIKHEF-K in the late 1980s and 1990s~\cite{Quint1,Quint2,Irene1,Irene2}. The available missing energy spectra\textemdash measured with a
resolution of  better than 100 KeV and extending up to $\sim 30$ MeV\textemdash provide both position and width of the peaks
corresponding to bound states of the recoiling $\isotope[207][]{Tl}$ nucleus.

\begin{figure}[h!]
\begin{center}
\includegraphics[scale=0.525]{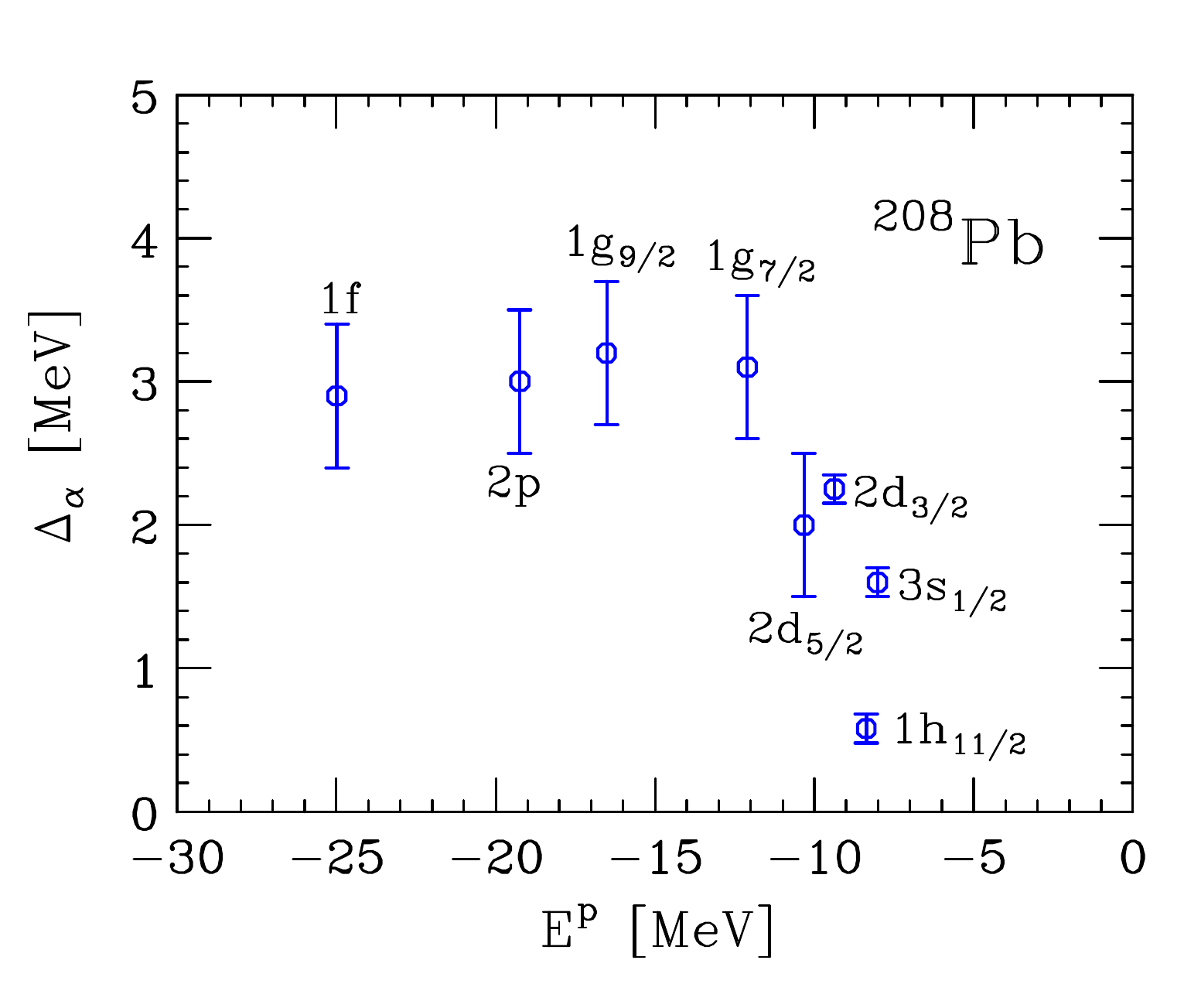}
\end{center}
\vspace*{-.2in}
\caption{Difference between the energies corresponding to the peaks of the missing energy
spectrum of the $\isotope[208][]{Pb}(e,e^\prime p)\isotope[207][]{Tl}$ reaction reported in Ref.~\cite{Quint1} and
the results of the mean-field calculations of Ref.~\cite{meanfield}, displayed as a function of the proton binding energy $E_p = -E^p_{\rm miss}$.
The states are labeled according to the standard spectroscopic notation}
\label{deviations}
\end{figure}

\begin{figure}[h!]
\begin{center}
\includegraphics[scale=0.525]{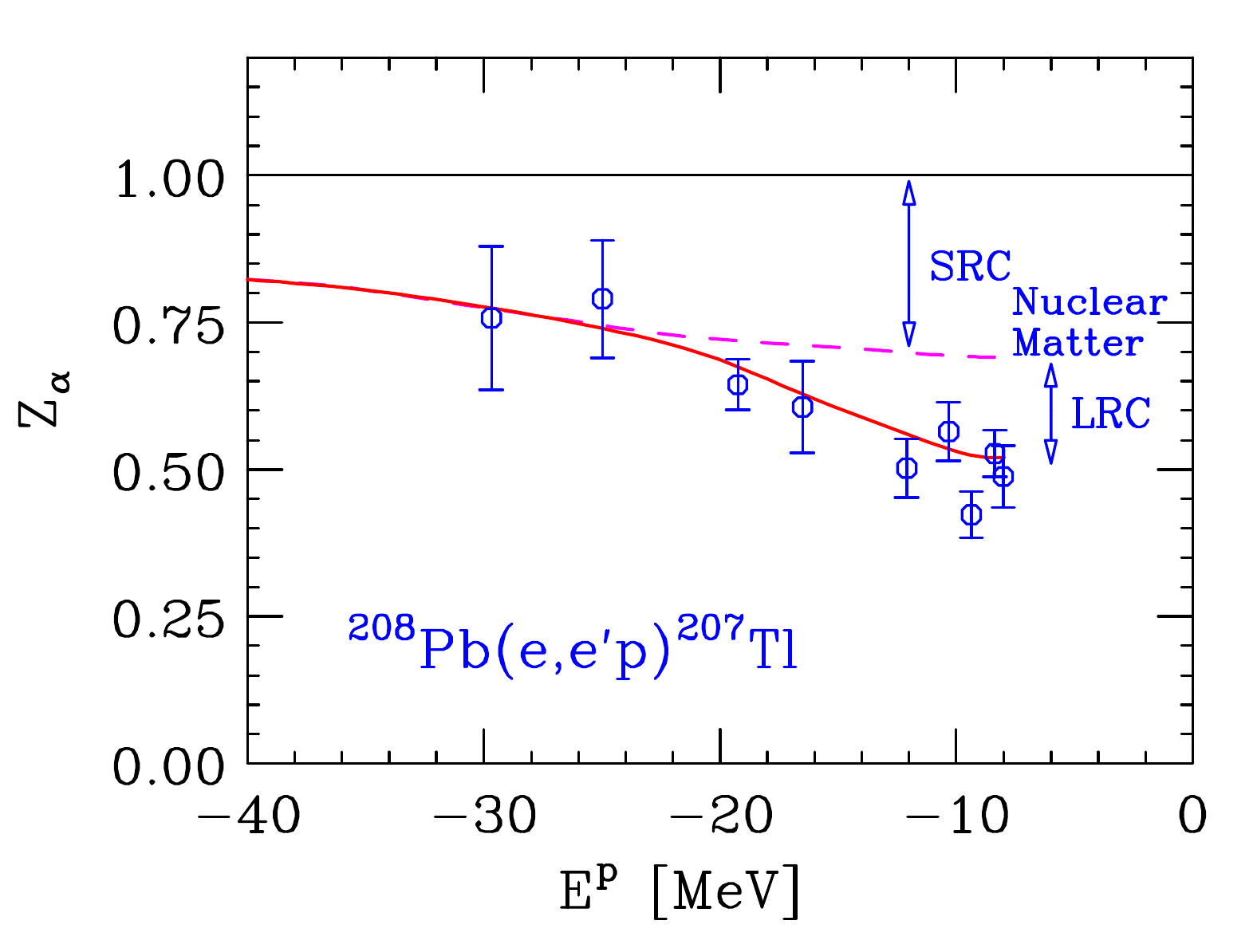}
\end{center}
\vspace*{-.2in}
\caption{Spectroscopic factors of the shell model states of  $\isotope[208][]{\rm Pb}$, obtained from the
analysis of the $\isotope[208][]{{\rm Pb}}(e,e^\prime p) \isotope[207][]{Tl}$ cross section measured
at NIKHEF-K~\cite{Quint1}. The dashed line represent the results of theoretical calculations of the spectroscopic factors
of nuclear matter, while the solid line has been obtained including corrections taking into account finite size and shell effects  in
$\isotope[208][]{\rm Pb}$~\cite{BFF0}. For comparison, the horizontal line shows the prediction of the
spectral function of Eq.~(\ref{SF:N:MF}). The deviations arising form short- and long-range correlations are highlighted, and labelled SRC and LRC, respectively.}
\label{spectrum}
\end{figure}

It is very important to realise that a meaningful interpretation of NIKHEF-K data requires the use of a theoretical framework
taking into account effects of nuclear dynamics beyond the mean-field approximation. This issue is clearly illustrated in Figs.~\ref{deviations}
and \ref{spectrum}.

Figure~\ref{deviations} displays the difference between the energies corresponding to the peaks in the measured missing energy
spectrum,  $\langle E^p_\alpha \rangle$, and the predictions of the mean-field model reported in Ref.~\cite{meanfield}, $E_\alpha^{MF}$.
It is apparent that the discrepancy, measured by the quantity
\begin{equation}
\label{def:delta}
\Delta_\alpha = | E_\alpha^{MF} - \langle E^p_\alpha \rangle | \ ,
\end{equation}
where the index $\alpha \equiv \{ nj\ell \}$ specifies the state of the recoiling system, is sizeable, and
  as large as $\sim 3$ MeV for deeply bound states.

In Fig.~\ref{spectrum}, the spectroscopic factors extracted from NIKHEF-K data are compared to the
results of the theoretical analysis of Ref.~\cite{BFF0}. The solid line, exhibiting a remarkable agreement with the experiment, has been
obtained combining theoretical nuclear matter results, displayed by the dashed line, and a phenomenological correction to the imaginary
part of the nucleon
self-energy, accounting for finite size and shell effects.
The energy dependence of the spectroscopic factors of nuclear matter at equilibrium density has been derived from a calculation of the
pole contribution to the spectral function of Eq.~(\ref{SF:N}), carried out using Correlated Basis Function (CBF) perturbation theory and a microscopic nuclear Hamiltonian including phenomenological two- and three-nucleon potentials~\cite{GF1}.

The results of Fig.~\ref{spectrum} show that the spectroscopic factors of $\isotope[208][]{\rm Pb}$,
defined as
\begin{equation}
Z_\alpha = \int \frac{d^3k}{(2\pi)^3} | \chi_\alpha ({\bf k})|^2 \ ,
\end{equation}
with $\chi_\alpha ({\bf k})$ being the Fourier transform of the overlap between the target ground state and the state of the recoiling nucleus, featuring
a hole in the shell model orbit $\alpha$. It appears that in the case of deeply bound proton states $Z_\alpha$ is largely
 determined by short-range correlations, moving strength to the continuum component of the spectral function. 
As a consequence, they are  largely unaffected by surface and shell effect, and can be accurately estimated using the results of nuclear matter calculations. Finite size effects, mainly driven by long-range nuclear dynamics, are more significant in the vicinity of the Fermi surface, where they account for up to $\sim$ 35\% of the deviation from the prediction obtained using the mean field spectral function of Eq.~(\ref{SF:N:MF}), represented by the solid horizontal line.

In addition to the nucleon spectral function, the analysis of the $\isotope[208][]{\rm Pb}(e, e^\prime K^+)\isotope[208][\Lambda]{\rm Tl}$
cross section requires a consistent description of the $\Lambda$ spectral function, defined by  Eq.~\eqref{SF:L}.
Following the pioneering nuclear matter study of Ref.~\cite{wim}, microscopic calculations of $P_\Lambda({\bf k}_\Lambda,E_\Lambda)$  in a variety of hypernuclei, ranging from
$\isotope[5][\Lambda]{\rm He}$ to  $\isotope[208][\Lambda]{\rm Pb}$, have been recently carried out by the author of  Ref.~\cite{Isaac}.
In this work, the self-energy of the $\Lambda$\textemdash simply related to the corresponding Green's function\textemdash was obtained from $G$-matrix perturbation theory in the Brueckner-Hartree-Fock approximation, using the J\"ulich~\cite{julich1,julich2} and Nijmegen~\cite{nijmegen1,nijmegen2,nijmegen3} models of the YN potential.

The generalisation of the approach of Ref.~\cite{Isaac}\textemdash needed to treat $\isotope[208][\Lambda]{\rm Tl}$ using Hamiltonians
including both YN and YNN potentials\textemdash does not appear to involve severe difficulties, of either conceptual or technical nature.
Therefore, a consistent
description of the $\isotope[208][]{\rm Pb}(e, e^\prime K^+)\isotope[208][\Lambda]{\rm Tl}$ process within the factorisation scheme derived in this work
is expected to be achievable within the time frame relevant to the JLab experimental program.

Different computational approaches, mostly based on the Monte Carlo method~\cite{Carlson:2014vla},
have been very successful in obtaining ground-state expectation values of Hamiltonians involving
nucleons and hyperons, needed to model the equation of state of strange baryon matter, see, e.g. Ref.~\cite{puzzle}. However, the present development of these techniques
does not allow the calculation of either $(e,e^\prime p)$ or $(e,e^\prime K^+)$ cross sections, most notably in the kinematical regime
in which the underlying non-relativistic approximation is no longer applicable. On the other hand, the
approach based on factorisation, allowing for a fully relativistic treatment of the electron-proton interaction, has proved very effective for the interpretation of the body of available  $(e,e^\prime p)$ data.

\section{Summary and outlook}
\label{summary}

The results discussed in this paper\textemdash in which kaon electro production from nuclei is analysed within the framework of many-body theory using, for the first time, the Green's function formalism\textemdash show that valuable new information on hypernuclear dynamics can be obtained from the
$\isotope[208][]{{\rm Pb}}(e,e^\prime K^+) \isotope[208][\Lambda]{Tl}$ cross section, the measurement of which in JLab Hall A has been recently approved~\cite{franco_AIP}.   In this context, an important role will be played by the
availability of accurate missing energy spectra measured by $\isotope[208][]{{\rm Pb}}(e,e^\prime p) \isotope[207][]{Tl}$ experiments, which can
be employed to obtain a largely model-independent determination of the $\Lambda$ binding energies
in $\isotope[208][\Lambda]{Tl}$.

Owing to the extended region of constant density, $\isotope[208][]{{\rm Pb}}$ is the best available proxy for uniform nuclear matter.
This feature, which also emerges from the results displayed in Fig.~\ref{spectrum}, will be critical to acquire new information on
three-body forces, complementary to that obtainable using a Calcium target.

The results of accurate many-body calculations of the ground-state energies of  finite nuclei~\cite{CVMC} and isospin-symmetric nuclear matter~\cite{APR}\textemdash performed with the {\em same} nuclear Hamiltonian, including the Argonne $v_{18}$~\cite{AV18} and
Urbana IX~\cite{UIX} NN and NNN interaction models, respectively\textemdash  show that the potential energy per nucleon arising from three-nucleon interactions is a monotonically increasing function of A, whose value changes sign, varying from  $-0.23$ MeV in $\isotope[40][]{{\rm Ca}}$ to 2.78 MeV in nuclear matter at equilibrium density.
Therefore, constraining three-body forces in the mass region in which their effect changes from attractive to repulsive in the non-strange sector appears to be needed, especially in view of astrophysical applications involving matter at supranuclear densities, in which three-body interactions are known to play a critical role.

The solution of the so called {\em hyperon puzzle} is likely to require a great deal of theoretical and experimental work
for many years to come. The extension of the JLab kaon
electro production program to $\isotope[208][]{{\rm Pb}}$ has the potential to provide an important contribution,
 needed to broaden the present understanding of hypernuclear dynamics in nuclear matt

\acknowledgments
This work was supported by the Italian National Institute for Nuclear Research (INFN) under grant TEONGRAV.
The author is deeply indebted to Petr Byd{\v z}ovsk{\'y}, Franco Garibaldi and Isaac Vida\~na for many 
illuminating discussions on issues related to the subject of this article. 

\end{document}